\documentclass{aa}
\usepackage{epsfig}
\usepackage{graphicx}
\begin{document}
\title{Absorption systems in the spectrum of GRB~021004
             \thanks{
Based on observations made with the Nordic Optical Telescope,
operated on the island of La Palma jointly by Denmark, Finland,
Iceland, Norway, and Sweden.} }

\author{P. M\o ller \inst{1}
        \and J.P.U. Fynbo \inst{2,3}
        \and J. Hjorth \inst{3}
	\and B. Thomsen \inst{2}
	\and M.P. Egholm  \inst{4,2}
	\and M.I. Andersen \inst{5}
	\and J. Gorosabel \inst{6,7}
	\and S.T. Holland \inst{8}
	\and P.~Jakobsson \inst{3}
        \and B.L. Jensen \inst{3}
	\and H. Pedersen \inst{3}
	\and K. Pedersen \inst{3}
	\and M. Weidinger \inst{2}
        }

\offprints{P. M\o ller}
\institute{
         European Southern Observatory, Karl-Schwarzschild-Stra\ss e 2,
         D-85748, Garching by M\"unchen, Germany
	 \and
	 Department of Physics and Astronomy, \AA rhus University, 
	 Ny Munkegade, DK-8000 \AA rhus C, Denmark
	 \and
	 Astronomical Observatory,
	 University of Copenhagen,
	 Juliane Maries Vej 30, DK--2100 Copenhagen \O, Denmark
         \and
	 Nordic Optical Telescope, Apartado Postal 474,
	 38700 Santa Cruz de La Palma, Canary Islands, Spain
         \and
	 Astrophysikalisches Institut Potsdam, An der Sternwarte 16, 
	 D-14482 Potsdam, Germany 
	 \and
	 Instituto de Astrof\'{\i}sica de Andaluc\'{\i}a (IAA-CSIC),
         P.O. Box 03004, E-18080 Granada, Spain
         \and
	 Laboratorio de Astrof\'{\i}sica Espacial y F\'{\i}sica
         Fundamental, P.O. Box 50727, E--28080 Madrid, Spain
         \and
	 Department of Physics, University of Notre Dame,
         Notre Dame, IN 46556-5670, U.S.A.
         }
\mail{pmoller@eso.org \& jfynbo@phys.au.dk}

\date{Received ; accepted }

\abstract{We report on a 3600 s spectrum of GRB 021004 obtained
with the Nordic Optical Telecope on La Palma 10.71 hours after the 
burst. We identify absorption lines from five systems at redshifts 1.3806,
1.6039, 2.2983, 2.3230, and 2.3292. In addition we find an emission
line which, if due to Ly$\alpha$ from the host galaxy, gives a 
redshift of 2.3351. The nearest absorber is blueshifted by 530 km
s$^{-1}$ with respect to this line, consistent with shifts seen in
Damped Ly$\alpha$ and Lyman-Break galaxies at similar redshifts. The 
emission line flux is $2.46\pm0.50 \times 10^{-16}$ erg s$^{-1}$ cm$^{-2}$.
Some of the absorption systems are ``line-locked'', an effect often
seen in QSO absorption systems believed to originate close to the QSO
central engine.
}

   \maketitle

\section{Introduction}

   More than 30 Optical Afterglows (OAs) to Gamma-Ray Bursts (GRBs) have
been detected to date. Absorption lines due to metal enriched gas in the host 
galaxy of the GRB have been detected in all cases where a spectrum with a
good signal-to-noise ratio of the OA has been obtained. Prior to GRB~021004
the redshift has been determined from such detected metal absorption for 
about a dozen OAs.
Absorption line spectra of GRBs have primarily been used to constrain the 
redshift of the burst but also to gain insight into the nature of the burst,
the gas phase of its host galaxy, and of its large scale environment 
(e.g. Vreeswijk et al. 2001; Mirabal et al. 2002).

Furthermore, it is interesting to consider how the GRB 
absorption line systems compare to the well-studied 
QSO absorption line systems, and to examine whether the physical
conditions (e.g. column densities, metallicities, ionizations states,
and kinematics) in the absorbers detected in GRB afterglow spectra are
similar to any of the various classes of QSO absorbers (e.g. Jensen et al. 
2001; Savaglio et al. 2002a; Salamanca et al. 2002).

In this {\it Letter} we present optical spectroscopy of the afterglow
of GRB~021004. 
GRB~021004 was detected by the HETE-II satellite on October 4.5043
2002 UT (Shirasaki et al. 2002). An optical afterglow 
was reported already 10 minutes after the burst (Fox 2002). The first 
optical spectroscopy reported two absorption systems at
z=1.38 and 1.60 (Fox et al. 2002), which
were later confirmed by Eracleous et al. (2002), Anupama et al. 
(2002) and Castander et al. (2002). Ly$\alpha$ absorption at z=2.323 
was first reported by Chornock \& Filippenko (2002).
The presence of this system was confirmed by Salamanca et al. 
(2002b), Savaglio et al. (2002b), Sahu et al. (2002) and 
Castro-Tirado et al. (2002).
Its lightcurve showed strong deviations from the
usual power-law decay indicating structure in the surrounding medium 
(Lazzati et al. 2002; Holland et al. in prep). 

\section{Observations and data reduction}

\begin{figure*}[ht]
\begin{center}
\epsfig{file=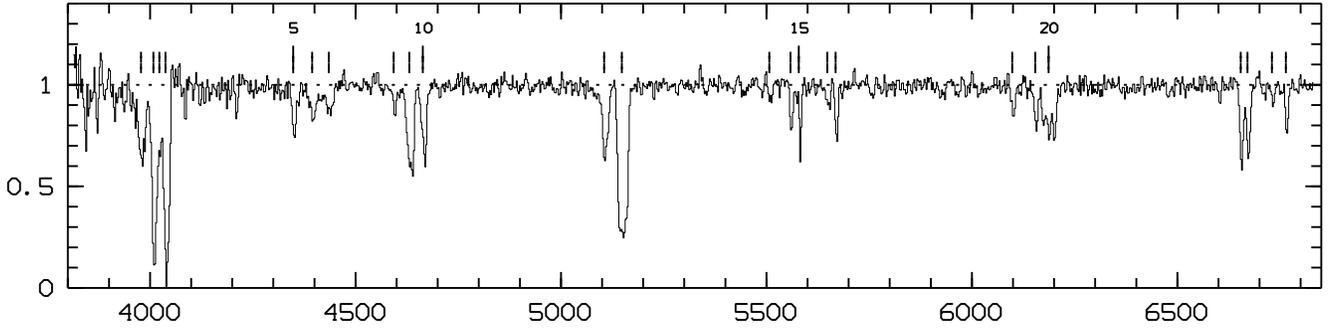,width=4.6cm,angle=270,clip=}
\caption{Normalised spectrum of GRB 021004. Absorption lines are
marked and numbered as in Table~1.}
\end{center}
\end{figure*}

\subsection{Observations}
The spectrum of GRB 021004 presented in Fig.~1 was obtained with the
2.56-m Nordic Optical Telescope (NOT) using the Andaluc\'{\i}a Faint
Object Spectrograph (ALFOSC). The detector was a 2048$^2$ pixels
thinned Loral CCD with a pixel scale of 0\farcs189. We used a grism 
covering 3850\AA -- 6850 \AA\  and a slit of 1.3 arcsec providing a 
resolution of 6.5 \AA\ and a sampling of 1.5 \AA\ per pixel.

The spectroscopic observations consisted of a single 3600 s exposure 
started at Oct. 4.9505 (10.71 hrs after the burst).
Photometric observations obtained on the same and the following nights
at the NOT and elsewhere made it possible to closely follow the
GRB light-curve. The brightness of the OA at the start of the
exposure was R=18.48$\pm$0.03 and B=19.15$\pm$0.10. During the
one-hour exposure the OA faded by 0.08 magnitudes. These broad band 
magnitudes were used for the flux calibration.

\subsection{Data reduction and absorption line-list}
Standard techniques were used for bias, dark and flat field corrections.
For the optimal 1D extraction, and for the line-search, measurement,
identification, and fitting we used a code originally developed for
2D spectral PSF fitting (M{\o}ller 2000) and for QSO spectral analysis
(for details see M{\o}ller \& Kj{\ae}rgaard 1992).

In Table~1 we list centroid (vacuum corrected) and observed
equivalent width of all absorption lines found above our 5$\sigma$
detection limit. Several of the listed lines are complex blends, and
for those we provide the total observed quantities with no attempt at
this point to deblend them. Inferred redshifts are listed only for
the single unblended lines. The absorption lines are marked above
the normalised spectrum shown in Fig.~1.

\begin{table}
\caption{Absorption lines in GRB 021004}
\begin{tabular}{rccclc}
\hline
No. & $\lambda_{\rm vac}$ & W$_{\rm obs}$ & $\sigma_{\rm W}$ & line ID
(system) & $z_{\rm abs}$ \\
    & (\AA ) & (\AA ) & (\AA ) & & \\
\hline
 1 & 3981.75 &  7.10 & 0.63 & complex (Fig.~2) & \\
 2 & 4011.90 & 13.11 & 0.54 & complex (Fig.~2) & \\
 3 & 4026.32 &  2.46 & 0.31 &   &  \\
 4 & 4041.43 & 14.18 & 0.47 & complex (Fig.~2) & \\
 5 & 4352.06 &  2.79 & 0.30 & \ion{Al}{ii} (B) & 1.6048 \\
 6 & 4398.21 &  2.95 & 0.35 & \ion{C}{ii} (C)  & 2.2957 \\
 7 & 4438.65 &  2.57 & 0.34 & \ion{C}{ii} (D+E) & \\
 8 & 4596.82 &  1.26 & 0.21 & \ion{Si}{iv} (C) & 2.2982 \\
 9 & 4634.89 &  7.97 & 0.29 & \ion{Si}{iv} (C+D+E) &  \\
10 & 4667.64 &  4.95 & 0.28 & \ion{Si}{iv} (D+E) &  \\
11 & 5108.88 &  5.96 & 0.26 & \ion{C}{iv} (C) & \\
12 & 5152.16 & 19.56 & 0.26 & \ion{C}{iv} (D+E) & \\
13 & 5511.21 &  0.84 & 0.20 & \ion{Al}{ii} (C) & 2.2986 \\
14 & 5562.56 &  1.99 & 0.19 & \ion{Al}{ii} (E) & 2.3293 \\
$^a$15 & 5582.76 &  2.41 & ? & \ion{Fe}{ii} (A) &  \\
16 & 5652.16 &  1.32 & 0.19 & \ion{Fe}{ii} (A) & 1.3804 \\
17 & 5672.26 &  2.38 & 0.17 & \ion{Fe}{ii} (A) & 1.3805 \\
18 & 6102.42 &  1.64 & 0.23 & \ion{Fe}{ii} (B) & 1.6032 \\
19 & 6158.50 &  2.41 & 0.23 & \ion{Fe}{ii} (A) & 1.3809 \\
20 & 6190.76 &  7.58 & 0.33 & complex (Fig.~2) & \\
21 & 6658.57 &  4.19 & 0.22 & \ion{Mg}{ii} (A) & 1.3812 \\
22 & 6675.02 &  3.83 & 0.21 & \ion{Mg}{ii} (A) & 1.3809 \\
$^b$23 & 6734.36 &  0.84 & ? & \ion{Fe}{ii} (B) &  \\
24 & 6768.53 &  2.00 & 0.18 & \ion{Fe}{ii} (B) & 1.6031 \\
\hline
\end{tabular}
$^a$ Line 15 is influenced by strong sky subtraction residuals.
\vskip -2pt

$^b$ Line 23 is strongly influenced by a cosmic.
\end{table}

We have identified five absorption systems at redshifts 1.3806, 1.6039,
2.2983, 2.3230, and 2.3292 and we shall in what follows refer to them
as systems A, B, C, D, and E. The identification of those systems was
carried out independently of the early GCNCs, but is seen to agree
well with the circulars summarised in Sect.~1.

For systems C, D, and E our spectrum covers Ly$\alpha$. No line
redwards of the Ly$\alpha$ line of system E is unidentified, and we
conclude that line No 4 in Table 1 marks the onset of the Lyman forest.
System E is therefore most likely associated with the GRB itself,
and could mark the redshift of the GRB host galaxy. There are however
a few caveats and we shall return to this question in Sect.~4.

Most of the absorption lines are blended, making it impossible to
obtain a good redshift directly from line centroiding. The redshifts
given above were therefore derived through line profile fitting. Detailed
fits to the blends are shown in Fig.~2, and the redshifts required to
fit each species are listed in Table 2.

\begin{table}
\caption{Absorption redshift constraints from line fitting}
\begin{tabular}{cccccc}
\hline
  & \ion{Fe}{II} & \ion{C}{iv} & \ion{Al}{ii} & \ion{Si}{iv} & Total \\
\hline
A & 1.3806 & -      & -      & -      & $1.3806\pm 0.0005$ \\
B & 1.6030 & -      & 1.6048 & -      & $1.6039\pm 0.0009$ \\
C & -      & 2.2980 & 2.2990 & 2.2982 & $2.2983\pm 0.0004$ \\
D & -      & 2.3220 &      - & 2.3240 & $2.3230\pm 0.0010$ \\
E & -      & 2.3290 & 2.3290 & 2.3296 & $2.3292\pm 0.0004$ \\
\hline
\end{tabular}
\end{table}

The lines are heavily saturated and the resolution low enough that the
line profiles are dominated by the resolution. Therefore the column
densities used in the fits are mostly too poorly constrained to be
useful. Nevertheless, from the Ly$\alpha$ line profiles we can
set a strict upper limit of $1.1\times 10^{20}$ cm$^{-2}$ on the 
\ion{H}{i} column density. The total \ion{H}{i} column actually used in 
the fit of systems C, D, and E was $5.7\times 10^{19}$ cm$^{-2}$.

\begin{figure}[ht]
\begin{center}
\epsfig{file=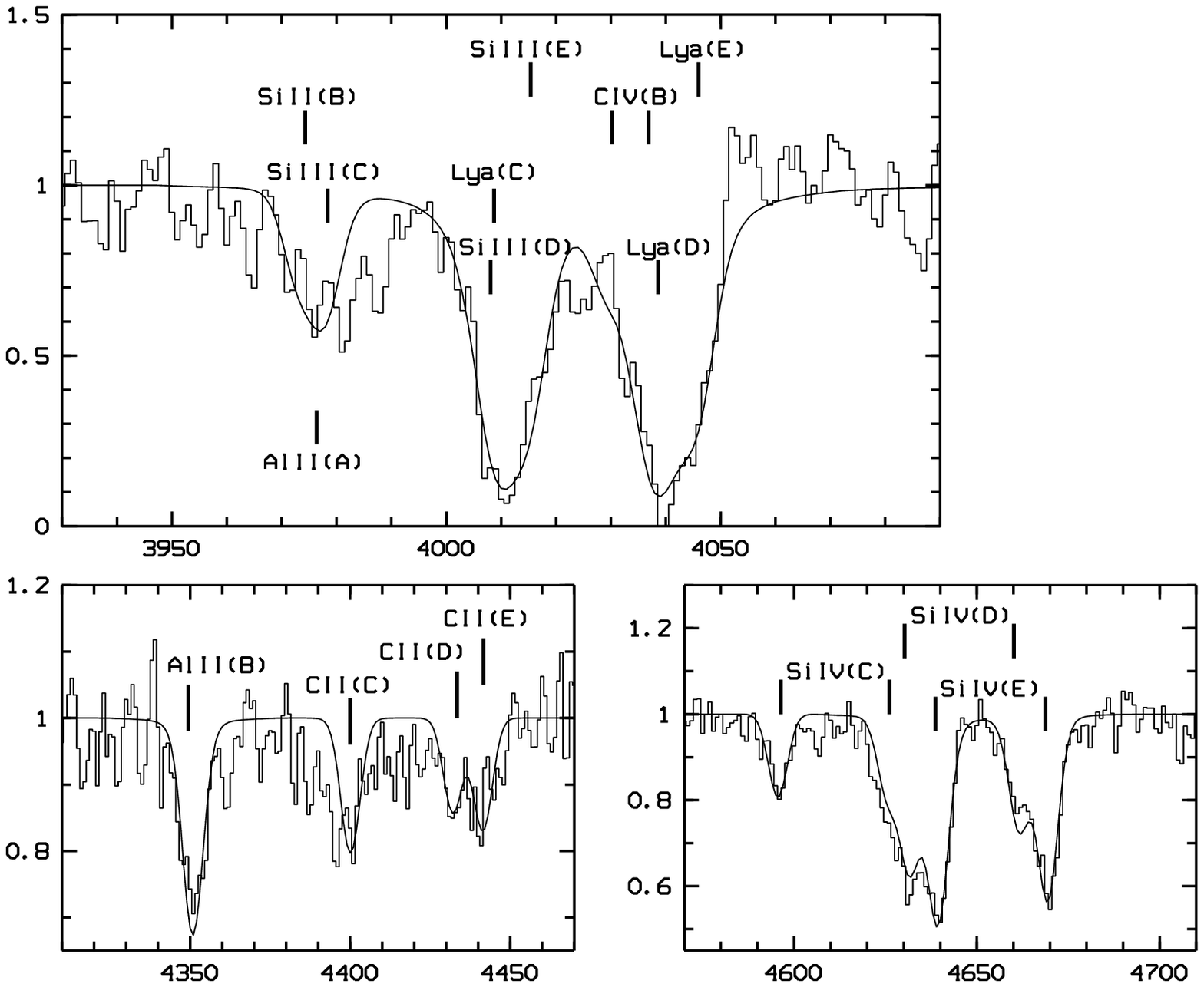,width=9cm,clip=}
\epsfig{file=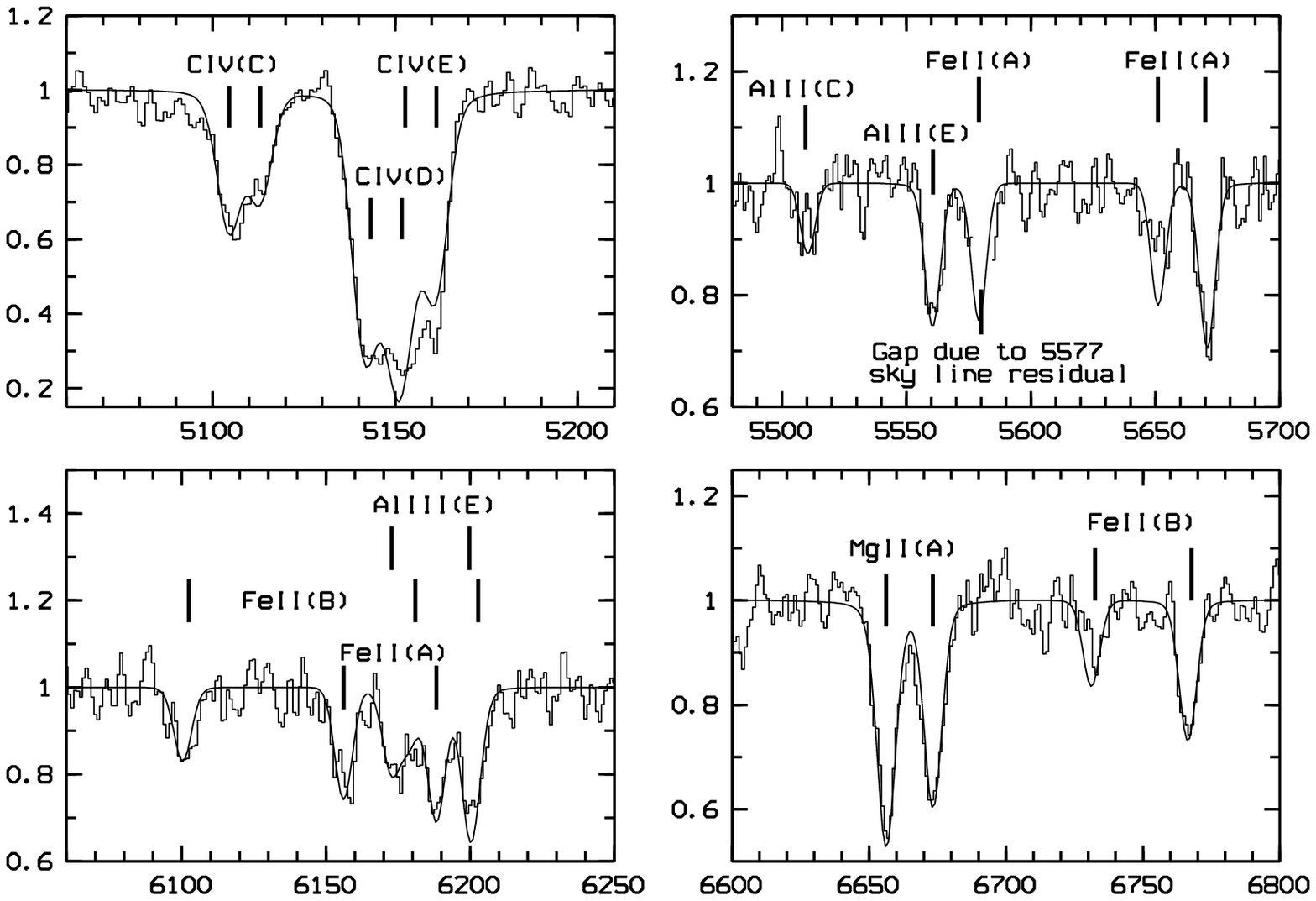,width=9cm,clip=}
\caption{Profile fits and line identifications.}
\end{center}
\end{figure}

\section{Ly$\alpha$ in emission}

Inspection of the top panel of Fig.~2 shows that what is mostly a
very good fit of the Ly$\alpha$ absorption lines of systems C, D, and
E, becomes very poor on the red shoulder of the last line.
Subtraction of the fit (Fig.~3) reveals that the poor fit is caused
by the presence of an emission line shortly redwards of the
Ly$\alpha$-E line. As suggested in several GCN Circulars
(Chornock \& Filippenko 2002; Salamanca et al. 2002b; Djorgovski
et al. 2002; Castro-Tirado et al. 2002), this line is
likely Ly$\alpha$ emission from the host galaxy. The line (Fig.~3,
right panel) is unresolved, has its centroid at 4054.4 \AA , and
a flux of $2.46\pm0.50 \times 10^{-16}$ erg s$^{-1}$ cm$^{-2}$
identical to other reported measurements to within $1 \sigma$.
The flux is comparable to the Ly$\alpha$ flux of the GRB~000926
host galaxy (Fynbo et al. 2002). The inferred redshift (2.3351) is 
530 km s$^{-1}$ higher than that of the E system. At the time of
writing the limit on the host galaxy magnitude is B$>$24, providing
a lower limit on the observed emission line equivalent width of
170 \AA .

\begin{figure}[ht]
\begin{center}
\epsfig{file=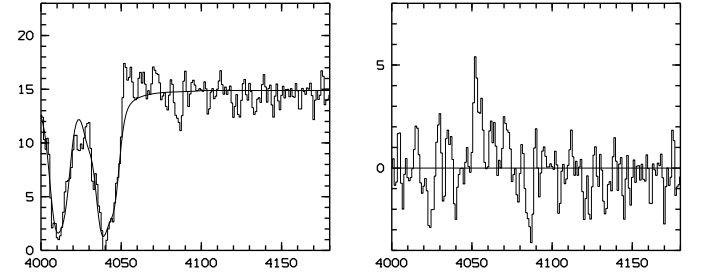,width=9cm,clip=}
\caption{Left: Section of the spectrum with Ly$\alpha$ absorption
lines. Note the poor fit to the extreme right wing of the second line.
Right: Residual after subtraction of the fit shown to the left. A
narrow Ly$\alpha$ emission line is now clearly visible. The vertical
scale in both plots is in units of $10^{-17}$ erg s$^{-1}$ cm$^{-2}$ \AA $^{-1}$.
}
\end{center}
\end{figure}

\section{Discussion}
\subsection{Properties of Ly$\alpha$ absorption and emission lines}
A shift of a few hundred km s$^{-1}$ between redshifts determined from
Ly$\alpha$ absorption and emission is commonly seen for DLA galaxies
(M{\o}ller et al.  2002) and for Lyman-Break galaxies (e.g. Adelberger 
at al. 2002, their Figures 2 and 3).
Two effects are thought to contribute to this observed velocity shift:
if the absorbing gas is in a galactic wind moving towards us, that will
blueshift the absorber and at the same time cause the resonantly
scattered Ly$\alpha$ photons to become redshifted in order to escape
the cloud. A spread in absorption/emission redshifts as large as
that seen in systems C--E and host Ly$\alpha$ emission
(3300 km s$^{-1}$) has never
been seen in Lyman-Break Galaxy spectra, and in fact equation 2 from
Adelberger et al. (2002) shows that the velocity shift of the
Ly$\alpha$ emission line should be less than 200 km s$^{-1}$ for an
emission line rest equivalent width larger than 50 \AA . Therefore
the large velocity range spanned here cannot be explained in terms
of normal host galaxy properties, and we conclude that the absorbers
must be in the local environment of the GRB.

\subsection{The nature of the GRB absorption systems}

Any absorption seen in the spectra of high-redshift QSOs may be
classified as belonging to one of three basic categories (Weymann et al.
1979).  The ``intervening systems'' are cosmologically distributed and 
can have any redshift lower than that of the source, ``local cluster
systems'' will have the same redshift as the source to within $\pm$ the
velocity dispersion of the local cluster or filament, and
``ejected systems'' (physically closest) can be radiatively
accelerated to large velocities.

In QSOs the ejected systems are often seen as BALs (P-Cygni type broad
absorption line systems) and/or as line-locked systems
(Foltz et al. 1987). Ejection velocities as high as 0.1c have been
reported (Vilkoviskij \& Irwin 2001), and the systems
are found to be highly ionized and to have extremely high metallicities 
(Savaglio et al. 1994; M{\o}ller et al. 1994). The high ionization is 
easily understood because of the intense UV flux of the QSO. The 
line-locking is explained as an effect of cloudlets being accelerated
via absorption at a given transition until its wavelength falls into the
shadow of another line in a cloud in front of it (Vilkoviskij et al.
1999).

In this scheme we would classify systems A and B as intervening systems,
while systems C, D and E display a surprising similarity to the
ejected systems of QSOs. They all have strong absorption of highly
ionized \ion{C}{IV} and \ion{Si}{IV} but no detectable \ion{Si}{II} 
absorption, they have
high column densities compared to stellar winds, and at intermediate
resolution they show evidence for line-locking in \ion{C}{IV},
\ion{Si}{III}/Ly$\alpha$ and possibly \ion{Si}{IV} (see also Savaglio 
et al. 2002b; Salamanca et al. 2002b). For confirmation of the line-locking
higher resolution spectra (already obtained) are needed.

It is difficult to understand why there is this resemblance between
systems C--E and ejected systems of QSOs.
We concluded above that the systems arise in the local environment of
the GRB. They may therefore be related either directly or indirectly
to the GRB or the GRB progenitor itself, or they may be caused by an
unrelated phenomenon which just happens to be sharing the same
volume of space. Firstly we would consider it unlikely that the
systems are directly related to the burst itself. So soon after the burst
any material ejected simultaneous with the GRB itself should undergo 
significant changes on relatively short timescales, yet there is no 
evidence for such changes between our spectrum and that of Matheson et 
al. 2002. 
Secondly one may ask if stellar winds could create such signatures.
Radiatively driven winds with terminal velocities of up to several
1000 km s$^{-1}$ are ubiquitous for very hot stars such as O stars
and WR stars (Lucy \& Abbott 1994; Kudritzki 2002) but these winds
cannot reach the required column densities. Also, presumably for the
same reason, actual line-locking has never been observed in stellar
winds.

We cannot exclude that the GRB is located close
to the inner region of a recently deceased QSO, especially at z$\sim$2,
where the QSO density is high. However, given the existing evidence
that GRBs seem to be related to the deaths of massive stars it is
more likely that the GRB progenitor was a massive, hot star and that
this star radiatively drove the fast outflow now observed in absorption
against the light of the afterglow. Other massive stars and supernova
explosions could be contributing to the wind if the progenitor was
located in a compact star-forming region similar to that of GRB~980425
(Fynbo et al. 2001). A similar scenario was suggested for
GRB~971214 by Ahn (2000).

Alternatively, as suggested by Lazzati et al. (2002), in the supra-nova 
scenario (Vietri \& Stella 1998) the wind and the clumpy surrounding
medium could be the result of a supernova predating the GRB by several
years.

\begin{acknowledgements}
We thank E. Y. Vilkoviskij for helpful discussions and our referee Davide 
Lazzati for fast, thorough and constructive criticism.
We are grateful to the staff at the Nordic Optical Telescope for keeping 
the Telescope working so well and for excellent support. JPUF gratefully
acknowledge support from the Carlsberg foundation. This work is supported 
by the Danish Natural Science Research Council (SNF).
\end{acknowledgements}


\begin{thebibliography}{}
\bibitem{ASS}Adelberger, K.L., Steidel C.C., Shapley, A.E. \& Pettini, M., 
2002, ApJ in press 
\bibitem{Ahn}Ahn, S.H., 2000, ApJL, 520, 9
\bibitem{India} Anupama, G.C., et al. 2002, GCNC \#1582
\bibitem{CPG}Castro-Tirado, A. J. et al. 2002, GCNC \#1635
\bibitem{CF} Chornock, R. \& Filippenko, A. V. 2002, GCNC \#1605
\bibitem{D2002}Djorgovski, S.G. et al. 2002, GCNC \#1620
\bibitem{ESM} Eracleous, M. et al. 2002 GCNC \#1579
\bibitem{Foltz} Foltz, C. B., Weymann, R. J., Morris, S. L. \& Turnshek,
D. A. 1987, ApJ, 317, 450
\bibitem{Fox} Fox, D. W., 2002, GCNC \#1564
\bibitem{F2} Fox, D. W. et al., 2002, GCNC \#1569
\bibitem{FHA2001} Fynbo, J. P. U., Holland, S.T., Andersen, M.I. et al., 2001,
ApJL, 542, L89
\bibitem{FMT2002} Fynbo, J. P. U., M{\o}ller, P., Thomsen, B. et al., 2002,
A\&A, 388, 42
\bibitem{JFG}Jensen, B.L., Fynbo, J.U., Gorosabal, J. et al., 2001, 
A\&A, 370, 909
\bibitem{K2002} Kudritzki, R.-P., 2002, ApJ, 577, 389 
\bibitem{L} Lazzati, D., et al., 2002, submitted to A\&A Letters
\bibitem{LA} Lucy, L.B., \& Abbott, D.C., 1993, ApJ, 405, 738
\bibitem{Mat} Matheson, T., Garnavich, P. M. et al. 2002 submitted to ApJ
\bibitem{MHK2002}Mirabal, N., Halpern, J.P., Kularni, S.R. et al.
2002, ApJ, 578, 818
\bibitem{M} M{\o}ller, P. 2000, {\it The Messenger}, 99, 31
\bibitem{MK} M{\o}ller, P. \& Kj{\ae}rgaard, P. 1992, A\&A, 258, 234
\bibitem{MJP94} M{\o}ller, P., Jakobsen, P., \& Perryman, M. A. C. 1994
A\&A, 287, 719
\bibitem{MWF2002} M{\o}ller, P., Warren, S. J., Fall., S. M., Fynbo, J. U., 
\& Jakobsen, P. 2002, ApJ, 574, 51
\bibitem{SFB} Sahu, K., et al. 2002, GCNC \#1608
\bibitem{SRW} Salamanca, I. et al. 2002a, submitted to MNRAS 
\bibitem{SRW2} Salamanca, I. et al. 2002b, GCNC \#1611
\bibitem{SS} Savaglio, S., D'Odorico, S. \& M{\o}ller, P. 1994,
A\&A, 281, 331
\bibitem{SFF} Savaglio, S., Fall, S.M., Fiore, F., submitted to ApJ
\bibitem{SS2} Savaglio, S. et al. 2002b, GCNC \#1633
\bibitem{SGM} Shirasaki, Y. et al. 2002, GCNC \#1565
\bibitem{VS1998} Vietri, M., \& Stella, L., 1998, ApJL, 507, L45
\bibitem{Vil01} Vilkoviskij, E. Y. \& Irwin, M. J. 2001, MNRAS, 321, 4
\bibitem{Vil99} Vilkoviskij, E. Y., Efimov, S. N., Karpova, O. G.,
\& Pavlova, L. A. 1999, MNRAS, 309, 80
\bibitem{VFK}Vreeswijk, P. M., Fruchter, A., Kaper, L. et al. 2001,
ApJ, 546, 672
\bibitem{WWPT} Weymann, R. J., Williams, R. E. Turnshek, et al.
1979, ApJ, 234
\end{thebibliography}
\end{document}